\documentclass[11pt]{article}
\usepackage{bm}% bold math
\usepackage{latexsym}
\begin{document}

%\markboth{Jun Goryo}{the polar Kerr effect in a chiral $p$-wave superconductor}

\title{Intrinsic and extrinsic origins of the polar Kerr effect in a chiral $p$-wave superconductor}
\author{Jun Goryo}
\maketitle  

\begin{center}
{\it Department of Physics, Nagoya University, Furo-Cho, Chikusa-Ku, Nagoya, 464-8602\footnote{Present address: Institute of Industrial Science, 
the University of Tokyo, Komaba 4-6-1, Tokyo, 153-8505, Japan \\ E-MAIL: jungoryo@iis.u-tokyo.ac.jp}}
\end{center}

\begin{abstract}
Recently, the measurement of the polar Kerr effect (PKE) in the quasi two-dimensional superconductor Sr$_2$RuO$_4$, which is motivated to observe the chirality of $p_x + i p_y$-wave pairing, has been reported. We clarify that the PKE has intrinsic and extrinsic (disorder-induced) origins. The extrinsic contribution would be dominant in the PKE experiment. 
\end{abstract}

\section{Introduction} 

The polar Kerr effect (PKE)  in which the direction of polarization of reflected linearly polarized light is rotated  
is renowned as a phenomenon of the light in ferromagnetic compounds\cite{LRO}.
Recent measurement on the PKE in the quasi two-dimensional superconductor Sr$_2$RuO$_4$\cite{Xia} is motivated 
to observe the chirality of $p_x + i p_y$-wave state, which 
is the most plausible pairing symmetry in this superconductor\cite{Mackenzie-Maeno}. 

We suppose that $z>0$ is empty and $z<0$ filled by the superconductor and incident polarized light propagates 
along $z$ direction. The Kerr rotation angle is given by \cite{LRO,Argyres} 
\begin{eqnarray}
\theta_K = -{\rm Im}\left[\frac{\omega (q_z^{+} - q_z^{-})}{\omega^2 - q_z^{+}q_z^{-}}\right].  
\label{Kerr}
\end{eqnarray}
where $q_z^{\pm}=\sqrt{\omega^2+i\omega \sigma_{xx}(\omega)\pm \omega\sigma_{xy}(\omega)}$, which is 
the dispersion relation of light inside the superconductor, and $\sigma_{xx}$ and $\sigma_{xy}$ are the diagonal and Hall parts of the conductivity tensor, respectively. It is well established that $\sigma_{xx}(\omega)$ in the high 
frequency regime $\omega >> 2 |\Delta|$ ($|\Delta|$; amplitude of the gap function) is given by the Drude form 
$\sigma_{xx}(\omega)=\omega_p^2 \tau_0(1-i \omega \tau_0)^{-1}$, where $\omega_p$ is the plasma frequency and $\tau_0$ quasiparticle life time\cite{Mattis-Bardeen,Katsufuji}. 

We see from the numerator of Eq. (\ref{Kerr}) that the Hall conductivity is crucial to the PKE. In the following sections we discuss intrinsic and extrinsic parts of the Hall conductivity and obtain the Kerr angle (the terminologies ``intrinsic" and ``extrinsic" has been originally used in the anomalous Hall effect and spin Hall effect \cite{intrinsic-extrinsic}). To obtain the Hall conductivity, we use the Kubo formula 
\begin{equation}
\sigma_{xy} (\omega, {\bf q}_{\perp}) = \frac{1}{2 i \omega} \epsilon_{ij} <J_i(\omega,{\bf q}_{\perp}) J_j(-\omega,-{\bf q}_{\perp})>, 
\label{Kubo}
\end{equation}
where ${\bf q}_{\perp}=(q_x,q_y)$ and ${\bf J}(\omega, {\bf q})=-e \sum_{\bf p}\frac{2 {\bf p} + {\bf q}_{\perp}}{2 m_e} \Psi^\dagger_{\bf p} \Psi_{{\bf p}+{\bf q}_{\perp}}$ is the electric current.

\section{Intrinsic Hall conductivity and PKE: Clean limit}
We consider the clean limit in this section. 
We start from a single band model which would describe the $\gamma$-band in Sr$_2$RuO$_4$. The Hamiltonian for the Nambu Fermion field $\Psi_{\bf p}=(c_{{\bf p}\uparrow}, c^*_{-{\bf p}\downarrow})^T$ is 
\begin{eqnarray}
\hat{H}=
\left(
\begin{array}{cc}
\epsilon ({\bf p} + e {\bf A}) & \Delta_{\bf p} \\
\Delta^*_{\bf p} & - \epsilon ({\bf p} - e {\bf A})
\end{array}
\right), 
\label{Hamiltonian}
\end{eqnarray}
where $\epsilon({\bf p})=\frac{p_x^2+p_y^2}{2 m_e} - \epsilon_F$ and $\Delta_{\bf p}=\Delta (\hat{p}_x + i \hat{p}_{y})$ .  We should note that, since the Cooper pair has the electric charge $-2e$, there is the phase field $\theta$ of the gap function ($U(1)$ Goldstone mode) $\Delta_{\bf p}=\tilde{\Delta}_{\bf p} e^{-2 i e \theta}$, where $\tilde{\Delta}_{\bf p}$ is the gauge invariant component of the gap function since the phase field restores the gauge invariance.  We can obtain the manifestly gauge-invariant current-current correlation by integrating out $\Psi_{\bf p}$ and also $\theta$\cite{Goryo-Ishikawa,Yakovenko,Roy-Kalline} (see also Appendix). Then, using Eq. (\ref{Kubo}), we have 
the intrinsic Hall conductivity 
%at $T=0$ and for small $\omega$ and ${\bf q}$\cite{Goryo-Ishikawa} 
\begin{eqnarray}
\sigma_{xy}^{(I)}(\omega,{\bf q}_{\perp})&=&a(\omega,{\bf q}_\perp)\frac{-v_F^2 {\bf q}_{\perp}^2}{\omega^2 - v_F^2 {\bf q}_{\perp}^2} 
\nonumber\\
a(\omega, {\bf q}_\perp)&=&\frac{e^2}{4 \pi d} N_w f^{(I)}(\omega, {\bf q}_{\perp}) , 
\label{intrinsic}
\end{eqnarray}
where $d \simeq 6.6 \AA$ is the inter-layer distance and $f^{(I)} (\omega, {\bf q}_{\perp})$ is a temperature-dependent regular function that satisfies $f^{(I)}(0,0)=1$\cite{Goryo-Sigrist,Holovitz-Golub,Yakovenko,Roy-Kalline}, and $\theta$-integral yields a singular factor $ \frac{-v_F^2 {\bf q}_{\perp}^2}{\omega^2 - v_F^2 {\bf q}_{\perp}^2}$.  The coefficient 
$$
N_w=\int \frac{d^2p}{4 \pi} \hat{\bf g}_{\bf p} \cdot ({\boldmath \partial} \hat{\bf g}_{\bf p} \times {\boldmath \partial} \hat{\bf g}_{\bf p})    
$$
is the winding number first suggested by Volovik\cite{Volovik} that equals to the chirality of the Cooper pair represented by the internal angular momentum $l_z$ (for the chiral $p$-wave state, $l_z=1$). See, also \cite{Goryo-Ishikawa,Read-Green,Furusaki-Matsumoto-Sigrist}. 

In the measurement \cite{Xia}, the incident beam was set up as ${\bf q}=\hat{{\bf z}} q_z$, then, $\sigma_{xy}^{(I)}(\omega, {\bf q}_{\perp})=0$. 
In the recent papers \cite{Yakovenko,Roy-Kalline},  the lens effect that focuses the light to a spot on the sample surface causes ${\bf q}_{\perp}\neq 0$ has been taken into account and found a finite $\sigma_{xy}(\omega, {\bf q}_{\perp})$, but the estimated Kerr angle is about 9 orders of magnitude smaller than the observed Kerr angle. 

It has been pointed out that an in-plane incident light could lead a rather large Kerr angle\cite{Goryo-Ishikawa2}. 

\section{Extrinsic Hall conductivity and PKE} 

We shall take into account impurity scattering and obtain the extrinsic Hall conductivity. In this section, we neglect the intrinsic contribution. By integrating out the Fermion field\cite{Goryo-Kerr} and the phase field $\theta$, we obtain the gauge-invariant current-current correlation function. By using the Kubo formula Eq. (\ref{Kubo}), the extrinsic part can be summarized in the general form (see Appendix), 
\begin{eqnarray}
\sigma_{xy}^{(E)} (\omega, {\bf q}_{\perp}) =b(\omega,{\bf q}_{\perp}) \frac{\omega^2}{\omega^2 - v_{\rm F}^2 {\bf q}_{\perp}^2}, 
\label{extrinsic}
\end{eqnarray}
where a singular factor  $\frac{\omega^2}{\omega^2 - v_{\rm F}^2 {\bf q}_{\perp}^2}$ comes from the integration of the phase field. 
In the experimental situation,  $\frac{\omega^2}{\omega^2 - v_{\rm F}^2 {\bf q}_{\perp}^2}\simeq 1$ and the extrinsic part would be dominant. 

The lowest-order contribution to the coefficient $b(\omega,{\bf q}_{\perp})$ is from the current-current correlation function 
with the vertex correction by a single delta-function type disorder within the second Born approximation, which is in the third order of the impurity strength.\cite{Goryo-Kerr}.  Although the diagram looks similar to the skew scattering diagram in the anomalous Hall effect in a ferromagnet\cite{Bruno}, the critical difference is that the scatterer we consider does not include the spin-orbit interaction. 

The extrinsic mechanism was first pointed out by Ref. \cite{Goryo-Kerr}. Recently, the frequency dependence of the result was corrected by Ref. \cite{Yakovenko2}. Additionally, a contribution from the ladder sum of the scatterings in the first Born approximation is investigated in detail\cite{Yakovenko2}. 

Although quantitative agreement with the experiment has not yet been obtained, we believe that the extrinsic mechanism would be the essence of the PKE in Sr$_2$RuO$_4$. More consideration would be needed for the estimation of $b(\omega, {\bf q})$. 

\section{Summary}

We have pointed out that  the PKE in a chiral $p$-wave superconductor 
has intrinsic and extrinsic (disorder-induced) origins. The general forms of these contributions has been clarified (See, Eqs. (\ref{intrinsic}) and (\ref{extrinsic}). See, also Eq. (\ref{Hall}) in Appendix). 
The intrinsic mechanism comes from the chirality of the chiral $p$-wave pairing state, and 
the extrinsic one comes from the combination of the pairing state and the delta-function type disorder. The latter would be the essence 
of the PKE observed in Sr$_2$RuO$_4$\cite{Xia}.

\section*{Acknowledgement}
The author is grateful to D. S. Hirashima, H. Kontani, Y. Maeno, and V. M. Yakovenko for useful discussions and comments. This work was supported by Grant-in-Aid for Scientific Research (No. 19740241) from the Ministry of Education, Culture, Sports, Science and Technology.

\section*{Appendix}
The aim of this appendix is to derive the manifestly gauge-invariant current-current correlation function and obtain the Hall conductivity by using Eq. (\ref{Kubo}). For simplicity, we consider the zero temperature. 
The extension to the finite temperature would be straightforward. First, we integrate out Fermion from Eq. (\ref{Hamiltonian}). The effective Lagrangian for gauge fields and the phase field (in the Gaussian approximation) is \cite{Goryo-Ishikawa,Yakovenko,Roy-Kalline,Volovik}
\begin{eqnarray}
{\cal{L}}_{eff}[A_\mu, \theta]&=&\frac{1}{2 \lambda_L(q)^2 v_F^2} (A_0(q) + i q_0 \theta(q)) (A_0(-q)-i q_0 \theta(-q)) 
\nonumber\\
&&- \frac{1}{2 \lambda_L(q)^2}  (A_i (q)+ i q_i \theta(q)) (A_i(-q)-i q_i \theta(-q)) 
\nonumber\\
&&-\frac{a(q)}{2} i \epsilon_{ij} ((A_0(q)+iq_0 \theta (q)) q_i A_j(-q) + A_i(q) q_j (A_0(-q)-iq_0 \theta(q))
\nonumber\\
&& + \frac{b(q)}{2} i \epsilon_{ij} A_i(q) q_0 A_j(-q), 
\label{eff-lagrangian}
\end{eqnarray}
where $q_\mu=(q_0,{\bf q}_{\perp})=(\omega,{\bf q}_{\perp})$. The first and second terms correspond to the coulomb screening term and Meissner term in the low-frequency and long-wavelength limit. The chiral $p$-wave pairing yields the third and the last terms. They are T-odd anomalous terms and in the low-frequency and long-wavelength limit, we refer to them as "Chern-Simons-like term". The last term vanishes in the clean limit\cite{Goryo-Ishikawa,Yakovenko,Roy-Kalline,Volovik,Read-Green,Furusaki-Matsumoto-Sigrist}, while becomes nonzero in a disordered system\cite{Goryo-Kerr}. Eq. (\ref{eff-lagrangian}) is in the gaussian form and we can easily integrate out the phase field $\theta$. After the integration, we have, 
\begin{equation}
\int {\cal{D}} \theta \exp(i {\cal{L}}_{eff}[A_\mu,\theta])=\exp(i {\cal{L}}^{\prime}_{eff}[A_\mu]), 
\end{equation}
where 
\begin{eqnarray}
{\cal{L}}^{\prime}_{eff}[A_\mu]=\frac{1}{2}A_\mu(q)\Pi_{\mu\nu}(q)A_\nu(-q), 
\end{eqnarray}
and 
\begin{eqnarray}
\Pi_{00}(q)&=&\frac{-v_F^2 {\bf q}_{\perp}^2}{\omega^2-v_F^2 {\bf q}_{\perp}^2}\frac{1}{\lambda_L(q)^2 v_F^2}, 
\\
\Pi_{0j}(q)&=&\Pi_{j0}(-q)
\nonumber\\
&=&-i a(q) \epsilon_{ij} q_i + \frac{\omega}{\omega^2 - v_F^2 {\bf q}_{\perp}^2}\left(\frac{q_j}{\lambda_L(q)^2} - i(a(q)-b(q)) \epsilon_{kj} \omega q_k \right), 
\\
\Pi_{ij}(q)&=&\frac{-1}{\lambda_L(q)^2} \delta_{ij} + i b(q) \epsilon_{ij} \omega 
\\
&&- \frac{\lambda_L(q)^2 v_F^2}{\omega^2-v_F^2{\bf q}_{\perp}^2}
\left(\frac{q_i }{\lambda_L(q)^2} 
-i (a(q)-b(q)) \epsilon_{ki} \omega q_k\right)\left(\frac{q_j}{\lambda_L(q)^2}  +i (a(q)-b(q)) \epsilon_{lj} \omega q_l\right). 
\nonumber
\end{eqnarray}
are the density-density, density-current, and current-current correlation functions, respectively. We see that they satisfy the condition for gauge invariance (charge conservation); $q_\mu \Pi_{\mu\nu}=\Pi_{\mu\nu} q_\nu=0$. From the Kubo formula Eq. (\ref{Kubo}), we obtain the Hall conductivity 
\begin{eqnarray}
\sigma_{xy}(\omega,{\bf q}_{\perp})&=&\frac{\epsilon_{ij}}{2 i \omega} \Pi_{ij}(\omega,{\bf q}_{\perp})
\nonumber\\
&=&\frac{-v_F^2 {\bf q}_{\perp}^2}{\omega^2 - v_F^2 {\bf q}_{\perp}^2} a(\omega,{\bf q}_{\perp}) + \frac{\omega^2}{\omega^2 - v_F^2 {\bf q}_{\perp}^2} b(\omega,{\bf q}_{\perp}), 
\label{Hall}
\end{eqnarray}
where the first term corresponds to the intrinsic part Eq. (\ref{intrinsic}) and the second term corresponds to the extrinsic part Eq. (\ref{extrinsic}). When $b(\omega,{\bf q}_{\perp})=0$, the result agrees with the literature discussing the clean limit\cite{Goryo-Ishikawa,Yakovenko,Roy-Kalline}. 

In the PKE measurement\cite{Xia}, ${\bf q}=\hat{\bf z} q_z$, and then, $\sigma_{xy}(\omega)=b(\omega)$. This fact indicates that the induction of the coefficient $b(\omega)$, namely, the disorder effect is crucial to the PKE measurement\cite{Goryo-Kerr}.

Finally, we write down ${\cal{L}}_{eff}^\prime[A_\mu]$ explicitly; 
\begin{eqnarray}
{\cal{L}}_{eff}^\prime[A_\mu]&=&\frac{1}{2 \lambda_L(q)^2 v_F^2} A_0(q)^TA_0(-q)^T- \frac{1}{2 \lambda_L(q)^2}A_i(q)^TA_i(-q)^T
\nonumber\\
&&-\frac{a(q)}{2}i\epsilon_{ij} (A_0^T(q)q_iA_j^T(q)+A_i^T(q)q_jA_0^T(-q)) 
\nonumber\\
&&+\frac{b(q)}{2} i \epsilon_{ij} A_i^T(q) q_0 A_j^T(-q), 
\end{eqnarray}
where, 
\begin{eqnarray}
A_0^T(q)&=&A_0(q) - \frac{i q_0}{q_0^2 - v_F^2 {\bf q}_\perp^2} (q_0 A_0 (q) - v_F^2 {\bf q}_\perp \cdot {\bf A}(q)), 
\\
A_i^T(q)&=&A_i(q) - \frac{i q_i}{q_0^2 - v_F^2 {\bf q}_\perp^2} (q_0 A_0 (q) - v_F^2 {\bf q}_\perp \cdot {\bf A}(q)), 
\end{eqnarray}
are the transversal components of gauge fields inside a media, namely, these are gauge invariant\cite{Goryo-Ishikawa}. 
 The current Eq. (1) in Ref. \cite{Goryo-Kerr} is obtained with a gauge fixing condition 
 $q_0 A_0 (q) - v_F^2 {\bf q}_\perp \cdot {\bf A}(q)=0$.

\end{document}